\documentclass[aps,prb,floatfix,twocolumn,preprintnumbers,amsmath,amssymb,groupedaddress,showpacs,showkeys,superscriptaddress]{revtex4-1}

\usepackage{graphicx}  
\usepackage{dcolumn}   
\usepackage{bm}        
\usepackage{color}     
\usepackage{amsmath,amssymb}
\usepackage{psfrag}
\usepackage{multirow}
\usepackage{bbold}
\usepackage{braket}  

\begin{document}
	\title{Chiral Majorana fermions in graphene from proximity-induced superconductivity}
	\author{Petra H\"{o}gl}
	\email[Emails to: ]{petra.hoegl@physik.uni-regensburg.de}
	\author{Tobias Frank}
	\author{Denis Kochan}
	\affiliation{Institute for Theoretical Physics, University of Regensburg, 93040 Regensburg, Germany}
	\author{Martin Gmitra}
	\affiliation{Department of Theoretical Physics and Astrophysics, Pavol Jozef \v{S}af\'{a}rik University, 04001 Ko\v{s}ice, Slovakia}
	\author{Jaroslav Fabian}
	\affiliation{Institute for Theoretical Physics, University of Regensburg, 93040 Regensburg, Germany}

	\begin{abstract}
		\vspace{0.05cm}
		We present a detailed theoretical study of chiral topological superconductor phases in proximity-superconducting graphene systems based on an effective model inspired by DFT simulations. Inducing s-wave superconductivity to quantum anomalous Hall effect systems leads to chiral topological superconductors. For out-of-plane magnetization we find topological superconducting phases with even numbers of chiral Majorana fermions per edge which is correlated to the opening of a nontrivial gap in the bulk system in the $\mathrm{ K}$-points and their connection under particle-hole symmetry. We show that in a quantum anomalous Hall insulator with in-plane magnetization and nontrivial gap opening at $\mathrm{M}$, the corresponding topological superconductor can be tuned to host only single chiral Majorana states at its edge which is promising for proposals exploiting such states for braiding operations. 
	\end{abstract}
	
	\maketitle

	\section{Introduction}	
	Topological superconductors are the natural hosts of Majorana fermions. In particle physics a Majorana fermion is a fundamental particle predicted by the real solution of the Dirac equation, which Ettore Majorana considered in 1937.\cite{Majorana1937}~The existence of such a particle, which is its own antiparticle and therefore must be charge neutral, is unsettled. A possible candidate is the neutrino.\cite{Elliot2015}
	
	In condensed matter physics Majorana fermions can occur as emergent quasiparticles in solids. These particles can be thought of as collective excitations of the quantum many-body state that describes the interacting electron system. To fulfill the condition of a Majorana fermion of being its own antiparticle it needs to be an equal superposition of electron and hole degrees of freedom. Coherent superpositions of electrons and holes naturally occur in superconductors. If such an equal superposition of electrons and holes exists exactly at zero energy it is a Majorana zero mode (MZM) that shows unique physical properties having no analog in high-energy physics.\cite{Elliot2015,Aguado2017}~The most prominent is the non-Abelian statistics with potential applications in topological quantum computation. Interchanging MZMs is called braiding, which is described by unitary transformations that form a braid group. This group is not rich enough to achieve universal quantum computation but combining it with unprotected operations enables fault tolerant quantum computation which is expected to still be much more robust against decoherence than a nontopological quantum computer.\cite{Nayak2008}
	
	Recently it was proposed that non-Abelian statistics can
	be realized also with chiral Majorana modes which appear as propagating edge states in 2D topological superconductors. A 2D chiral topological superconductor is the natural superconducting analog of a quantum anomalous Hall effect (QAHE) insulator.\cite{Qi2009,Schnyder2008,Qi2010}~It has a bulk band gap and chiral one-dimensional Majorana fermions at the edges.\cite{Volovik1988,Read2000}~When superconductivity is introduced to a QAHE system with Chern number $C$, it can be turned into a chiral topological superconductor with Bogoliubov-de Gennes (BdG) Chern number $C_{BdG}=2C$, where $C$ indicates the number of chiral fermionic edge states in QAHE and $C_{BdG}$ counts the number of chiral Majorana edge states reflecting the doubling of degrees of freedom when one describes superconducting systems by means of the BdG formalism. On the other hand this makes the $C_{BdG}=\pm1$ state particularly interesting as it means that at the edge of this system only a single chiral Majorana fermion propagates, which has half of the degrees of freedom of a fermionic edge state. This state can be achieved by tuning the system parameters and it forms the minimal possible topological state in 2D.\cite{Qi2010}
	
	The search for chiral Majorana fermions has been driven by their potential applications for topological quantum computation. It has been predicted that in a vortex of a chiral topological superconductor with $C_{BdG}=1$ a single MZM appears which gives rise to non-Abelian statistics.\cite{Volovik1999,Read2000,Ivanov2001}~A proposal to use the propagation of single chiral Majorana fermions with purely electrical manipulations (instead of bound MZMs) to implement topologically protected quantum gates on the mesoscopic scale has been developed.\cite{Lian2018}~Recently, another scheme that exploits the chiral motion along the edge of a topological superconductor to realize non-Abelian braid operations has been proposed.\cite{Beenakker2019}~The experimental realization of a single chiral Majorana fermion in a magnetically doped topological insulator in proximity to a superconductor could not be uniquely demonstrated.\cite{He2017,Ji2018,Kayyalha2019}
	
	In this paper, we investigate an alternative host material for chiral Majorana fermions, namely graphene, which provides a 2D platform whose electronic properties can be remarkably changed by proximity effects. The formation of chiral Majorana fermions in graphene is based on exploiting proximity effects which allow to find QAHE phases in graphene and also induce superconductivity when it is in a heterostructure with a 2D superconductor such as NbSe$_2$. 
	
	In particular, we study a symmetry-based tight-binding model for graphene which contains proximity induced spin-orbit coupling (SOC) and magnetization\cite{Gmitra2016,Frank2018,Hoegl2019}~with additionally taking into account superconductivity. For out-of-plane magnetization we find topological superconductor phases with even BdG Chern numbers and trace the pairwise appearance of edge states in the valleys back to the particle-hole symmetry that connects the Dirac points. We raise the question whether a BdG Chern number of 1 is possible in graphene based on proximity-induced superconductivity in QAHE phases. Starting from a QAHE phase introduced from in-plane magnetization we indeed find such a unique topological phase. In the corresponding topological superconductor we show sublattice asymmetry in the exchange coupling or in the superconducting pairing to lead to the desired single chiral Majorana fermion state. In quantum Hall graphene with external magnetic field, topological superconductivity has also been proposed.\cite{SanJose2015}
	
	\begin{figure}[b]
		\centering
		\includegraphics[width=0.75\columnwidth]{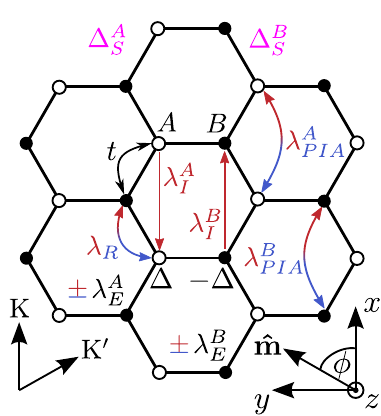}
		\caption{Scheme of graphene lattice with proximity-induced hoppings. Sublattice $A$ and $B$ 
			is denoted by empty and full dots, respectively. Color indicates action on spin (red spin up, blue spin down). The minimal model contains spin neutral 
			nearest-neighbor hopping $t$ and on-site staggered potential $\Delta$; spin-mixing nearest-neighbor 
			Rashba SOC $\lambda_R$; spin and sublattice resolved next-nearest neighbor intrinsic 
			SOC $\lambda^A_I$, $\lambda^B_I$; spin-mixing sublattice-resolved next-nearest neighbor PIA SOC $\lambda_{PIA}^A$, $\lambda_{PIA}^B$; on-site sublattice resolved exchange 
			splitting $\lambda^A_E$, $\lambda^B_E$ (spin-dependent energy shift spin up $+$, spin down $-$). The superconducting pairing is indicated by $\Delta_S^A$ and $\Delta_S^B$, assumed to have in general different amplitudes on $A$ and $B$ sublattices. The orientation of the reciprocal lattice is shown by $\mathrm{K}$ and $\mathrm{K'}$. The magnetization orientation in 
			real space is specified by $\mathbf{\hat{m}}$.}
		\label{fig:scheme_SC_model}
	\end{figure}
	The paper is organized as follows. In Sec. II, we introduce a tight-binding model and Bogoliubov-de Gennes Hamiltonian for superconducting graphene. In Sec. III, we use this to find nontrivial phases of the topological superconductor system for out-of-plane magnetization and corresponding edge states in zigzag and armchair nanoribbons. We investigate first a normal conducting system with in-plane magnetization in Sec. IV, which we use to find a topological superconductor with single chiral Majorana states per edge. In Sec. V, we summarize our results.

	\section{Tight-binding model and Bogoliubov-de Gennes Hamiltonian}\label{TB_BdG}
	We aim to investigate Majorana fermions which appear as self-conjugated, massless chiral edge states in 2D topological superconductors, where the topological superconductor is formed out of a QAHE system with proximity-induced superconductivity. We apply an effective tight-binding Hamiltonian for magnetic graphene\cite{Gmitra2016,Frank2018,Kochan2017,Hoegl2019}~and extend it by taking into account the formation of Cooper pairs in graphene due to proximity-induced s-wave superconductivity. The real-space tight-binding Hamiltonian is given by
	\begin{eqnarray}\label{eq:hamiltonian_SC}
	\mathcal{H}_S &=&
	-t\sum_{\left<i,j\right>,\sigma} c_{i\sigma}^\dagger c^{\phantom\dagger}_{j\sigma}+
	\Delta\sum_{i,\sigma} \, \xi_i \,c_{i\sigma}^\dagger c^{\phantom\dagger}_{i\sigma} \nonumber \\
	& &+\frac{2i\lambda_R}{3}\sum_{\left<i,j\right>,\sigma,\sigma^\prime}c_{i\sigma}^\dagger 
	c^{\phantom\dagger}_{j\sigma^\prime}\left[\left(\hat{\mathbf{s}}\times 
	\mathbf{d}_{ij}\right)_z\right]_{\sigma\sigma^\prime}\nonumber\\
	& &+\frac{i}{3}\sum_{\left<\left<i,j\right>\right>,\sigma,\sigma^\prime}c_{i\sigma}^\dagger 
	c^{\phantom\dagger}_{j\sigma^\prime}\left[\frac{\lambda_{ I}^{i}}{\sqrt{3}}\nu_{ij}\, \hat{s}_z \right.\nonumber\\
	&&+\left. 2\lambda_{PIA}^i\left(\hat{\sigma}_z\right)_{ij}\left(\mathbf{\hat{s}}\times\mathbf{D}_{ij}\right)_z^{\phantom\dagger} \right]_{\sigma\sigma^\prime}\nonumber \\
	& &+\sum_{i,\sigma,\sigma^\prime}\lambda_E^{i}\,c_{i\sigma}^\dagger c^{\phantom\dagger}_{i\sigma^\prime} 
	\left[\mathbf{\hat{m}}\cdot\mathbf{\hat{s}}\right]_{\sigma\sigma^\prime}\nonumber \\
	& &+\sum_{i,\sigma}\Delta^i_{S}\left(c_{i,\sigma}^\dagger c_{i,-\sigma}^\dagger + c^{\phantom\dagger}_{i,\sigma} c^{\phantom\dagger}_{i,-\sigma}\right),
	\end{eqnarray}
	where $c_{i\sigma}^\dagger\left(c_{i\sigma}^{\phantom\dagger}\right)$ is the creation (annihilation) operator for an 
	electron on lattice site $i$ that belongs to the sublattice $A$ or $B$ and carries spin $\sigma=\uparrow,\downarrow$. 
	
	The hoppings are indicated in Fig.~\ref{fig:scheme_SC_model}. The orbital terms of the QAHE part of the model are the nearest-neighbor hopping $t$ (sum over $\left<i,j\right>$) and the staggered on-site potential $\Delta$ ($\xi_{i}=\pm1$ on sublattice $A$/$B$). The spin-orbit part consists of the Rashba SOC $\lambda_R$, where the unit vector $\mathbf{d}_{ij}$ points from site $j$ 
	to $i$ and $\mathbf{\hat{s}}$ contains spin Pauli matrices, and the two sublattice-resolved next-nearest neighbor SOC terms (sum over 
	$\left<\left<i,j\right>\right>$) intrinsic SOC $\lambda^i_I$ and pseudospin inversion asymmetry (PIA) SOC $\lambda_{PIA}^i$ ($i=A,B$). Intrinsic SOC depends on clockwise ($\nu_{ij}=-1$) or counterclockwise ($\nu_{ij}=1$) 
	hopping paths from site $j$ to $i$. The PIA term contains $\mathbf{D}_{ij}$ the next-nearest neighbor unit vector pointing from site $j$ to $i$ and $\hat{\sigma}_z$ the pseudospin Pauli matrix. Time-reversal symmetry is broken by the magnetic part, the sublattice-resolved exchange coupling $\lambda_E^i$ ($i=A,B$). The orientation of magnetization is along the unit vector 
	$\mathbf{\hat{m}}=\left(\cos\phi\sin\theta,\sin\phi\sin\theta,\cos\theta\right)$, where $\phi$ is measured with respect to the $x$-axis and $\theta$ with respect to the $z$-axis in Fig.~\ref{fig:scheme_SC_model}. The last term introduces superconductivity to the QAHE model. The superconducting pairing is an on-site term that couples two electrons (or holes) with opposite spin. Due to the broken pseudospin symmetry we allow the superconducting pairing to be different on $A$ and $B$ sublattice $\Delta_S^A$, $\Delta_S^B$ (assumed to be real). Such a different proximity effect on the sublattices appears because the atoms in the graphene layer locally feel a different environment from the proximitized layers that often can be well described by an effective model with different values for the $A$ and $B$ sublattice. Energies are measured from the chemical potential which we set to zero throughout the paper. It can be controlled by gating or doping.
	
	To calculate bulk spectra of the superconducting system we need to transform the real-space tight-binding Hamiltonian to the BdG Hamiltonian in $\mathbf{k}$-space. For this we define the particle-hole symmetric Nambu spinor
	\begin{eqnarray}
	\boldsymbol{\Phi}_\mathbf{k}&=&\left[A^{\phantom\dagger}_\uparrow(\mathbf{k}),A^{\phantom\dagger}_\downarrow(\mathbf{k}),B^{\phantom\dagger}_\uparrow(\mathbf{k}),B^{\phantom\dagger}_\downarrow(\mathbf{k}),\right. \\ &&\left.A^{\dagger}_\downarrow(-\mathbf{k}),-A^{\dagger}_\uparrow(-\mathbf{k}),B^{\dagger}_\downarrow(-\mathbf{k}),-B^{\dagger}_\uparrow(-\mathbf{k}) \right]^T, \nonumber
	\end{eqnarray}
	containing creation (annihilation) operators $A_\sigma(\mathbf{k})$ ($A^\dagger_\sigma(\mathbf{k})$) for sublattice $A$ (and $B$), as basis which doubles the number of degrees of freedom of the system. For a given momentum $\mathbf{k}$ they are particle-hole, pseudospin, and spin. The Hamiltonian can be written as
	\begin{eqnarray}
	\mathcal{H}_S=\frac{1}{2}\sum_\mathbf{k}\,\boldsymbol{\Phi}_\mathbf{k}^\dagger H_{BdG}(\mathbf{k})\,\boldsymbol{\Phi}_\mathbf{k},
	\end{eqnarray}
	with the $8\times 8$ BdG Hamiltonian in $\mathbf{k}$-space $H_{BdG}(\mathbf{k})$ that describes the quasiparticle spectrum by
	\begin{eqnarray}
	H_{BdG}(\mathbf{k})\boldsymbol{\psi}_\mathbf{k}^n = E^n_\mathbf{k}\boldsymbol{\psi}^n_\mathbf{k},
	\end{eqnarray}
	with energy $E^n_\mathbf{k}$ of the $n$-th quasiparticle eigenstate
	\begin{eqnarray}
	\boldsymbol{\psi}^n_\mathbf{k}&=&\left[u^{A,n}_{\mathbf{k},\uparrow}, u^{A,n}_{\mathbf{k},\downarrow}, u^{B,n}_{\mathbf{k},\uparrow}, u^{B,n}_{\mathbf{k},\downarrow},\right. \\
	&& \left. (v^{A,n}_{-\mathbf{k},\downarrow})^\ast, -(v^{A,n}_{-\mathbf{k},\uparrow})^\ast, (v^{B,n}_{-\mathbf{k},\downarrow})^\ast, -(v^{B,n}_{-\mathbf{k},\uparrow})^\ast    \right]^T \nonumber
	\end{eqnarray}
	at momentum $\mathbf{k}$. The electron-like components are denoted by $u^{i,n}_{\mathbf{k},\sigma}$ and the hole-like components by $v^{i,n}_{\mathbf{k},\sigma}$ with spin $\sigma=\uparrow,\downarrow$ and sublattice $i=A,B$. With the basis defined above we get the BdG Hamiltonian in the form
	\begin{equation}
	H_{BdG}(\mathbf{k})=
	\left(
	\begin{array}{cc}
	H_e(\mathbf{k})                     &
	\Delta_S                                    \\
	\Delta_S                         &
	H_h(-\mathbf{k})
	\end{array}
	\right)\,,
	\label{eq:BdG_hamiltonian}
	\end{equation}
	where the electron-like Hamiltonian $H_e(\mathbf{k})$ is found by a Fourier transformation of the tight-binding model without superconductivity in Eq.~\ref{eq:hamiltonian_SC}. The hole-like Hamiltonian is obtained from the transformation
	\begin{eqnarray}
	H_h(\mathbf{k})=-\hat{\mathcal{T}}^{-1}H_e(\mathbf{k})\hat{\mathcal{T}}
	\end{eqnarray}
	with the antiunitary time-reversal symmetry operator $\hat{\mathcal{T}}=i\hat{\sigma}_0\hat{s}_y\hat{\mathcal{K}}$, where $\hat{\sigma}_i$ acts on pseudospin, $\hat{s}_i$ on spin-space, and $\hat{\mathcal{K}}$ denotes the complex conjugation operator. The sublattice-resolved superconducting pairing is given by
	\begin{eqnarray}
	\Delta_S=(\Delta^A_S\sigma_+-\Delta^B_S\sigma_-)s_0,
	\end{eqnarray}
	with the pseudospin matrices $\sigma_\pm=(\sigma_z\pm\sigma_0)/2$.
	
	The BdG Hamiltonian obeys --- by construction --- particle-hole symmetry, i.e.,
	\begin{eqnarray}
	\hat{\mathcal{P}}H_{BdG}(\mathbf{k})\hat{\mathcal{P}}^{-1} = -H_{BdG}(-\mathbf{k})
	\end{eqnarray}
	with the antiunitary particle-hole operator $\hat{\mathcal{P}}=\hat{\tau}_y\hat{\sigma}_0\hat{s}_y\hat{\mathcal{K}}$, where $\hat{\tau}_y$ is a Pauli matrix in particle-hole space and $\hat{\mathcal{P}}^2=1$. If we act with $\hat{\mathcal{P}}$ on a quasiparticle eigenstate we get
	\begin{eqnarray}
	\hat{\mathcal{P}}\boldsymbol{\psi}^n_\mathbf{k}&=&\left[v^{A,n}_{\mathbf{k},\uparrow}, v^{A,n}_{\mathbf{k},\downarrow}, v^{B,n}_{\mathbf{k},\uparrow}, v^{B,n}_{\mathbf{k},\downarrow},\right. \\
	&&\left.(u^{A,n}_{-\mathbf{k},\downarrow})^\ast, -(u^{A,n}_{-\mathbf{k},\uparrow})^\ast, (u^{B,n}_{-\mathbf{k},\downarrow})^\ast, -(u^{B,n}_{-\mathbf{k},\uparrow})^\ast    \right]^T. \nonumber
	\end{eqnarray}
	The transformation of the BdG Hamiltonian and its eigenstates under the particle-hole symmetry operator shows that it connects states with positive energy at $\mathbf{k}$ to states with negative energy at $-\mathbf{k}$. The Bogoliubov quasiparticle operator has the form
	\begin{eqnarray}
	\hat{\gamma}_n = \sum_\mathbf{k} \left(\boldsymbol{\psi}_\mathbf{k}^n\right)^\dagger \boldsymbol{\Phi}_\mathbf{k}.
	\end{eqnarray}
	It fulfills $\hat{\mathcal{P}}\hat{\gamma}_n=\hat{\gamma}_{-n}$, where the index $-n$ indicates that it is transformed to negative energy, pointing to the connection to localized MZMs, for which this is the Majorana condition for zero energy.
	
	In the following, we discuss two different configurations of this model.
	
	\section{Chiral Majorana fermions from quantum anomalous Hall effect with out-of-plane magnetization}\label{chiralMF_QAHE_oop}
	We investigate the effect of proximity-induced superconductivity on QAHE phases in graphene from uniform and staggered intrinsic spin-orbit and exchange couplings. Therefore, we consider the tight-binding Hamiltonian defined in Eq.~(\ref{eq:hamiltonian_SC}) with uniform and staggered intrinsic spin-orbit and exchange coupling, Rashba SOC, and staggered potential. We neglect for simplicity PIA SOC here. Only in the out-of-plane configuration of the magnetization the system exhibits nontrivial QAHE phases.\cite{Hoegl2019}~Therefore, we use out-of-plane orientation of magnetization. We focus on the two most interesting cases, namely the combination of staggered intrinsic SOC and uniform exchange coupling, which we refer to as (su), and uniform intrinsic SOC and staggered exchange, denoted (us). The type of the proximity-induced couplings in graphene -- staggered or uniform -- is determined by the proximitized materials. Graphene on transition-metal dichalcogenides (TMDCs) allows for staggered intrinsic SOC,\cite{Gmitra2015,Gmitra2016,Wang2015,Yang2016,Wang2016,Volkl2017,Zihlmann2018,Ghiasi2017, Cummings2017, Benitez2018}~whereas on a topological insulator the intrinsic SOC in graphene can be uniform.\cite{Song2018}~Similarly, the induced exchange coupling can be uniform\cite{Zollner2016,Predrag2016, Wang2015a,Leutenantsmeyer2017,Mendes2015,Swartz2012,Wei2016,Haugen2008,Yang2013,Hallal2017,Dyrdal2017,Zollner2018} or staggered as recently proposed.\cite{Hoegl2019}~ In this part we fix the superconducting pairing to be equal on $A$ and $B$,  $\Delta_S^A=\Delta_S^B=\Delta_S=0.03t$ and use generic model parameters $\Delta=0.1t$, $\lambda_R=0.075t$, $\lambda^A_E=|\lambda^B_E|=0.25t$ and 
	$\lambda^A_I=|\lambda^B_I|=3\sqrt{3}\cdot 0.06t$, if not indicated differently, to illustrate our findings.

	\subsection{Bulk band structure and phase diagrams}
\begin{figure}[t]
		\centering
		\includegraphics[width=1\columnwidth]{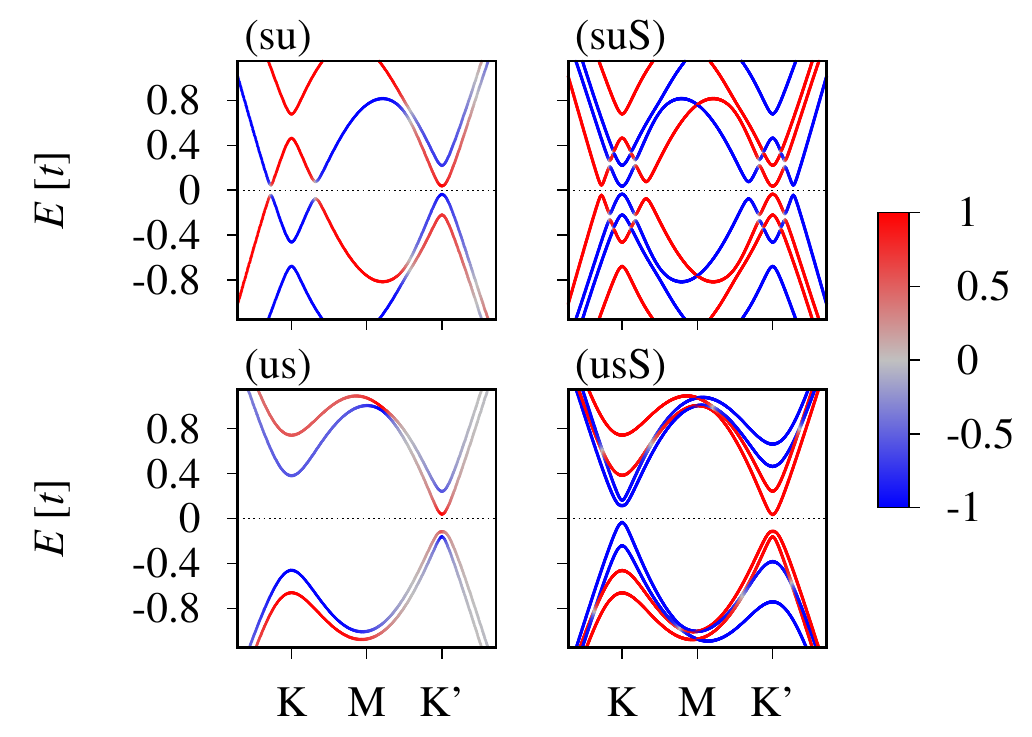}
		\caption{Calculated bulk band structure along $k_y=0$ for (su) $\lambda^A_I=-\lambda^B_I$, $\lambda^A_E=\lambda^B_E$, $\Delta_S=0$ and (suS) $\lambda^A_I=-\lambda^B_I$, $\lambda^A_E=\lambda^B_E$, $\Delta_S=0.03t$; (us ) $\lambda^A_I=\lambda^B_I$, $\lambda^A_E=-\lambda^B_E$, $\Delta_S=0$ and (usS) $\lambda^A_I=\lambda^B_I$, $\lambda^A_E=-\lambda^B_E$, $\Delta_S=0.03t$.
			In the left column color indicates spin expectation value  $\langle\hat{s}_z\rangle$ (red spin up, blue spin down) and in the right column it shows particle (red) or hole (blue) character $\langle \hat{\tau} \rangle_\mathbf{k}^n$ as defined in Eq.~(\ref{eq:particle-hole}). We use $\lambda^A_I=|\lambda^B_I|=3\sqrt{3}\cdot 0.06t$, $\lambda^A_E=|\lambda^B_E|=0.25t$, $\Delta=0.1t$, and $\lambda_R=0.075t$.}
		\label{fig:bulk_exchange_Z}
	\end{figure}

	We first compute bulk band structures from the BdG Hamiltonian in Eq.~(\ref{eq:BdG_hamiltonian}). As a consequence of the additional hole degree of freedom, we expect the number of bands in the superconducting spectra to double compared to the normal ones, which we calculate from the electron part of the Hamiltonian $H_e(\mathbf{ k})$ and indicate by $\Delta_S=0$. In Fig.~\ref{fig:bulk_exchange_Z} we compare the normal and superconducting spectrum for the two cases (su) and (us).
	
	To identify the particle or hole character of the superconducting bands we define a particle-hole expectation value as
	\begin{eqnarray}
	\langle \hat{\tau} \rangle_\mathbf{k}^n=\sum_{i\sigma} \left(\left| u^{i,n}_{\mathbf{k},\sigma}  \right|^2 -  \left| v^{i,n}_{-\mathbf{k},\sigma}  \right|^2 \right),
	\label{eq:particle-hole}
	\end{eqnarray}
	which sums up the probabilities of the components of the $n$-th eigenstate over sublattices $i=A,B$ and spin $\sigma$ and weights particle (hole) contributions positive (negative). In Fig.~\ref{fig:bulk_exchange_Z} we can trace back the appearance of additional hole bands compared to the normal system by mirroring the normal bands around the $E=0$ axis and an axis through the M-point (which is like transforming $\mathbf{k}\rightarrow -\mathbf{k}$). The superconducting coupling then induces gap openings. The superconducting spectra (suS) and (usS) exhibit (as constructed) particle-hole symmetry.
	
	To study whether the systems in the panels in Fig.~\ref{fig:bulk_exchange_Z} for (suS) and (usS) belong to topological superconducting phases we explore the bulk band gap and BdG Chern number in the $\lambda_I$-$\lambda_E$ parameter space. The topological character is determined by a Chern number since the 2D model breaks time-reversal and chiral symmetry but is particle-hole symmetric and thus, belongs to the symmetry class D of the periodic table of topological invariants.\cite{Ryu2010}~ The BdG Chern number is defined as
	\begin{eqnarray}
	C_{BdG}=\frac{1}{2\pi}\sum_m^{occ}\int_{BZ}d^2\mathbf{k}\,\Omega^m_{BdG,z}(\mathbf{k})
	\label{eq:BdG_chern}
	\end{eqnarray}
	where $\Omega^m_{BdG,z}(\mathbf{k})$ is the $z$-component of the Berry curvature for the $m$-th band in momentum space
	\begin{eqnarray}
	&\Omega^m_{BdG,z}(\mathbf{k})= \\
	\sum_{n \neq m} &\frac{-2\mathrm{Im} \left<\boldsymbol{\psi}_\mathbf{k}^m\right. \left|\partial_{k_x} H_{BdG}(\mathbf{k}) 
		\right|\left.\boldsymbol{\psi}_\mathbf{k}^n\right> \!\left<\boldsymbol{\psi}_\mathbf{k}^n\right.\left| \partial_{k_y} 
		H_{BdG}(\mathbf{k}) \right| \left.\boldsymbol{\psi}_\mathbf{k}^m\right> }{(E^m_{\mathbf{k}} -E^n_{\mathbf{k}} )^2}.\nonumber
	\label{eq:BdG_berry}
	\end{eqnarray}
	The BdG Chern number is calculated by means of the BdG Hamiltonian $H_{BdG}(\mathbf{k})$ [Eq.~(\ref{eq:BdG_hamiltonian})] with the quasiparticle wavefunctions 
	$\boldsymbol{\psi}_\mathbf{k}^m$ and eigenenergies $E^m_{\mathbf{k}}$. Summation runs over all occupied bands ($m=1,2,3,4$) and the integration is over the whole
	Brillouin zone.
	
\begin{figure}[t]
		\centering
		\includegraphics[width=1\columnwidth]{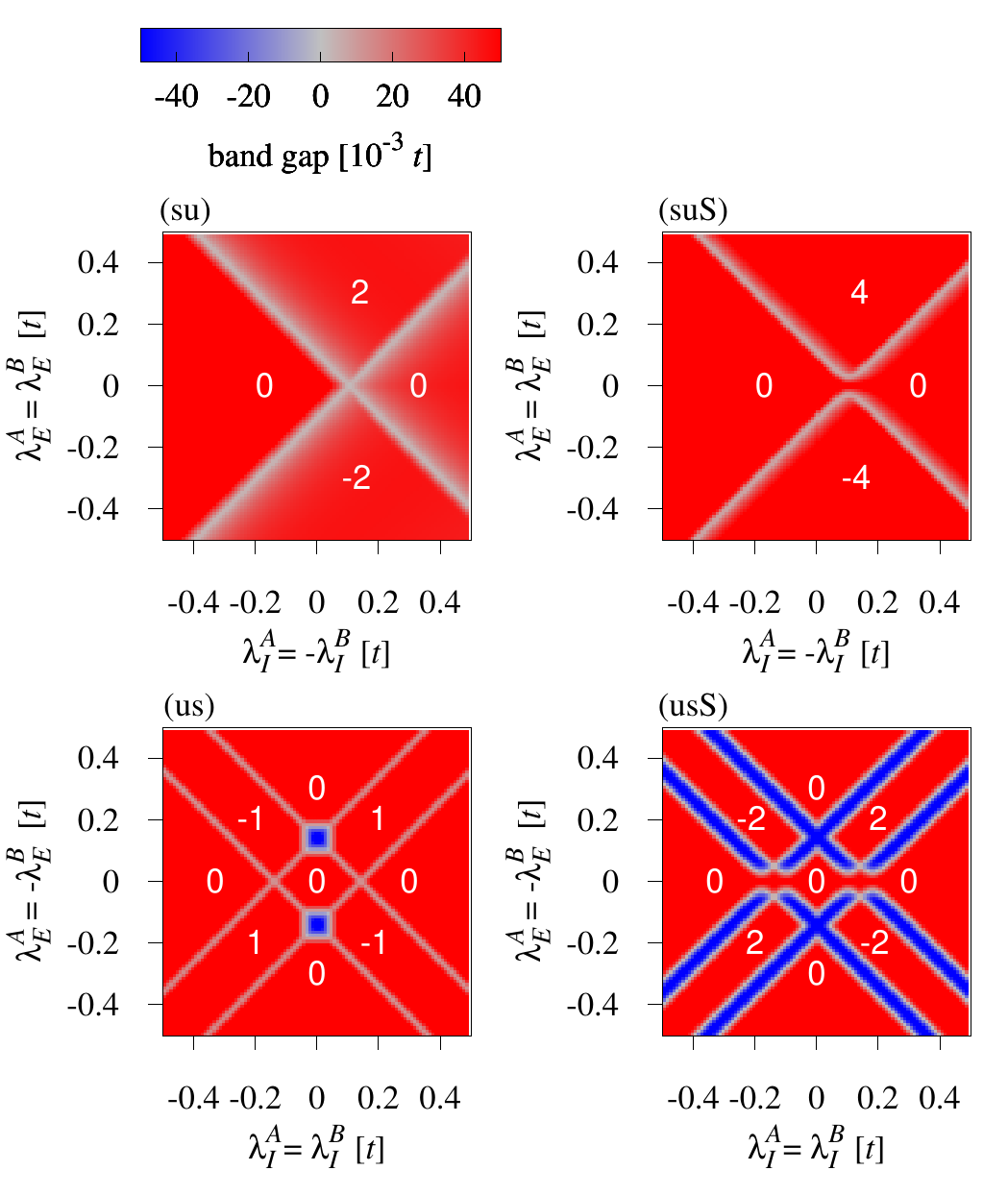}
		\caption{Topological phase diagrams for $\Delta_S=0$ in (su), (us) and $\Delta_S=0.03t$ in (suS), (usS). Global bulk band gap and (BdG) Chern 
			number (white numbers) for $\Delta=0.1t$ and 
			$\lambda_R=0.075t$ with varying intrinsic SOC $\lambda^A_I$, $\lambda^B_I$ and exchange 
			splitting $\lambda^A_E$, $\lambda^B_E$ for out-of-plane magnetization 
			$\mathbf{\hat{m}}=\left(0,0,1\right)$. Negative band gap indicates a transition to a metallic system due to 
			indirect band gap closing from bands at different $\mathbf{k}$ values.}
		\label{fig:gap_exchange_v3_minus_su_comps_norm_sc}
	\end{figure}

	The topological phase diagrams are presented in Fig.~\ref{fig:gap_exchange_v3_minus_su_comps_norm_sc}, where we compare the QAHE systems to their superconducting analogs. We expect to find regions (away from phase transitions) in phase space with Chern numbers $C_{BdG}=2C$ since in the BdG formalism we describe the quasiparticle degrees of freedom (electron and hole). Therefore, there is one copy of QAHE for each quasiparticle type. Indeed, we find this relation between the QAHE systems and topological chiral superconductors. For the (su) case we go from $C=\pm2$ to $C_{BdG}=\pm4$ and for the (us) case from $C=\pm1$ to $C_{BdG}=\pm2$. We determine analytical conditions to be in a topological superconductor phase from the gap closing condition for particle-hole symmetric systems $\det[H_{BdG}(\mathbf{k})]=0$ at $\mathrm{K}$ and $\mathrm{K'}$. The topological regions are confined by
	\begin{eqnarray}
	\mathrm{(suS)}\,\, |\lambda_E|&>&\sqrt{\Delta_S^2+(\lambda_I-\Delta)^2},\label{eq:bdgcondition1}\\
	\mathrm{(usS)}\,\,  |\lambda_E|&>&  \sqrt{\Delta_S^2+(\lambda_I-\Delta)^2}     \label{eq:bdgcondition2}\\
	\wedge \,\,|\lambda_E|&<&  \sqrt{\Delta_S^2+(\lambda_I+\Delta)^2}   \nonumber \\
	\wedge \,\,|\lambda_E|&>&  \sqrt{\Delta_S^2+\lambda_I^2+4\lambda_R^2+\Delta^2-2\sqrt{\lambda_I^2(4\lambda_R^2+\Delta^2)}} ,  \nonumber
	\end{eqnarray}
	with $|\lambda_I|=|\lambda^A_I|=|\lambda^B_I|$ and $|\lambda_E|=|\lambda^A_E|=|\lambda^B_E|$. The results agree exactly with the numerically found nontrivial regions in Figs.~\ref{fig:gap_exchange_v3_minus_su_comps_norm_sc}(suS) and (usS). For case (suS) the superconducting gap introduces an anticrossing to the topological phase boundaries and leads to symmetric band gap closing at $\mathrm{K}$ and $\mathrm{K'}$. For case (usS) we find four expressions (for each $\lambda_E\gtrless0$) for zero energy solutions but we need only three to define the border of the topological regions in Eq.~(\ref{eq:bdgcondition2}). The gap is closed by an indirect closing that happens simultaneously between one band in $\mathrm{K}$ and one in $\mathrm{K'}$ such that $E_{\mathrm{K/K'}}=0$. Before a trivial gap opens, the system first enters a metallic phase. Around $\lambda_E\approx0$, superconductivity induces anticrossings between the gap closing lines.

	\subsection{Chiral Majorana fermions in nanoribbons}
	For a chiral topological superconductor the BdG Chern number indicates the number of edge states in a finite system, counting chiral Majorana fermions, analogous to the Chern number for QAHE, specifying the number of chiral fermions. A chiral Majorana fermion has half of the degrees of freedom of a chiral fermionic edge state in the QAHE. Therefore, the BdG Chern number cannot be directly connected to a quantized Hall conductance that is proportional to the number of edge states. In fact, the transport properties of samples with chiral Majorana fermions and their unique signatures are a subject of ongoing research.\cite{He2017,Ji2018,Kayyalha2019}
	\begin{figure}[t]
		\centering
		\includegraphics[width=1\columnwidth]{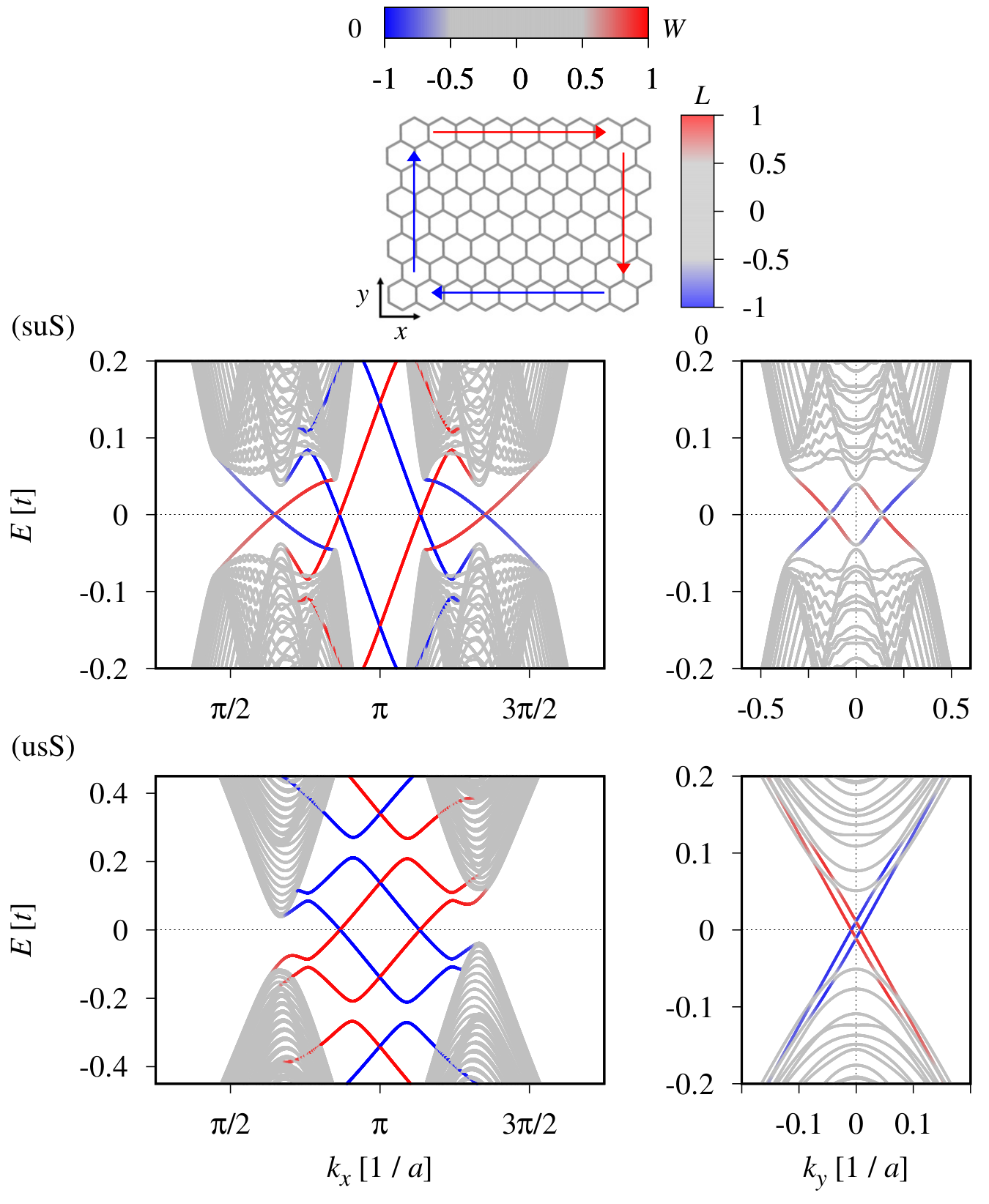}
		\caption{Calculated band structure of zigzag (left) and armchair (right) nanoribbons with  width $L$ ($W$) of 100 unit cells for zigzag (armchair). Color indicates localization of states as defined in Eq.~(\ref{eq:supp1}). The position and propagation direction of the edge states is shown in the scheme. (suS) shows $\lambda^A_I=-\lambda^B_I$, $\lambda^A_E=\lambda^B_E$ and (usS) $\lambda^A_I=\lambda^B_I$, $\lambda^A_E=-\lambda^B_E$ both with $\Delta_S=0.03t$, $\lambda^A_E=|\lambda^B_E|=0.25t$, $\lambda^A_I=|\lambda^B_I|=3\sqrt{3}\cdot 0.06t$, $\Delta=0.1t$, and $\lambda_R=0.075t$.}
		\label{fig:bands_100_loc_Sc}
	\end{figure} 
	
	To show the bulk-edge correspondence for chiral superconductors we compute band structures of zigzag and armchair nanoribbons presented in Fig.~\ref{fig:bands_100_loc_Sc}. For case (suS) we find four chiral Majorana states per edge in zigzag and armchair ribbons as expected from the BdG Chern number. In the zigzag ribbon they form pairs of valley-centered and intervalley states. In the armchair ribbon the states are degenerate and less localized. The case (usS) exhibits two well localized chiral states per edge in the zigzag as well as in the armchair terminated ribbon.
	
	To quantify the localization of the edge states in a nanoribbon, we sum the probability amplitude in the $i$-th 
	unit cell of the $m$-th quasiparticle eigenstate $\chi^m_i(k)$ weighted by the position of the unit cell in the ribbon 
	\begin{eqnarray}
	\sum_{i=0}^N (i-\frac{N}{2})|\chi_i^m(k)|^2 / \frac{N}{2},
	\label{eq:supp1}
	\end{eqnarray}
	where $N$ is the number of unit cells, $k$ is the momentum ($k=k_x$ for zigzag, $k=k_y$ for armchair ribbon), and we sum the electron and hole contributions. From the resulting 
	number that varies between $-1$ and $1$ we learn on which side of the ribbon a state propagates and how strong it is confined to the edge. 
	
	We have found only even numbers of chiral Majorana fermions for the systems investigated above. Can we tune our graphene systems to achieve the minimal topological state in 2D that shows one chiral Majorana fermion per edge? Due to the particle-hole symmetry, which connects states at $+E^n_{+\mathbf{ k}}$ to states at $-E^n_{-\mathbf{ k}}$, gap closings appear always simultaneously at $\mathrm{ K}$ and $\mathrm{K'}$ in our system. Therefore, the BdG Chern number changes always by $C_{BdG}=\pm2$ and one pair of chiral Majorana fermions localized on opposite edges is created symmetrically around each $\mathrm{K}$-point, i.e., we always have an even number of states per edge in this configuration.

	\section{Single chiral Majorana fermion from quantum anomalous Hall effect with in-plane magnetization}\label{chiralMF_QAHE_ip}
	We aim to find a chiral topological superconductor phase with $C_{BdG}=\pm1$. As we have discussed this is not possible in graphene when the topologically nontrivial gaps are formed in the valleys. Instead, such a gap should be introduced around the $\Gamma$- or $\mathrm{M}$-points. There are proposals for chiral topological superconductor systems in graphene where the Majorana fermions are created from a nontrivial gap at the $\Gamma$-point but very high chemical potential is necessary and the band gaps are very small since the bulk gap comes from the proximity-induced superconductivity only.\cite{Dutreix2014,Wang2016MF}
	
	\begin{figure}[t]
		\centering
		\includegraphics[width=1\columnwidth]{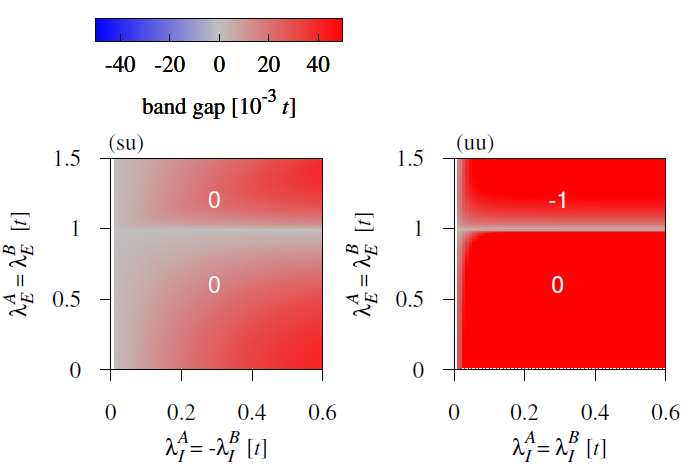}
		\caption{Global bulk band gap and Chern number (white numbers) for $\lambda_{PIA}=0.045t$ with varying intrinsic SOC $\lambda^A_I$, $\lambda^B_I$ and exchange 
			splitting $\lambda^A_E$, $\lambda^B_E$ for in-plane magnetization along 
			$\mathbf{\hat{m}}=\left(\sqrt{3}/2,1/2,0\right)$.}
		\label{fig:gap_ip_su_uu}
	\end{figure}
	We propose to use a QAHE phase with $C=1$ induced by a nontrivial gap at an $\mathrm{M}$-point\cite{Ren2016a,Ren2017}~and add proximity s-wave superconductivity to it. We consider the real-space Hamiltonian from Eq.~(\ref{eq:hamiltonian_SC}), taking into account intrinsic SOC and exchange coupling as before and additionally PIA SOC. We set Rashba SOC to zero, as too large $\lambda_R$ is detrimental for the QAHE phase (see Fig.~\ref{fig:gap_rashba_pia} in Appendix~\ref{appendix_rashba} for the discussion on the effects of the Rashba parameter) and we wish to limit the parameter space to the bare minimum to remove the unnecessary complexity and still see the desired effect. To this end we also neglect the staggered potential $\Delta$. We consider sublattice resolved terms for intrinsic spin-orbit and exchange coupling and now also for the superconducting pairing $\Delta_S^A$ and $\Delta_S^B$. We fix the numerical values for PIA SOC $\lambda_{ PIA}^A=\lambda_{ PIA}^B=\lambda_{PIA}=0.045t$ and the orientation of magnetization to $\theta=\pi/2$ and $\phi=\pi/6$ to have nondegenerate bands for in-plane magnetization at the $\mathrm{M}$-points. 
	
	\begin{figure}[t]
		\centering
		\includegraphics[width=1\columnwidth]{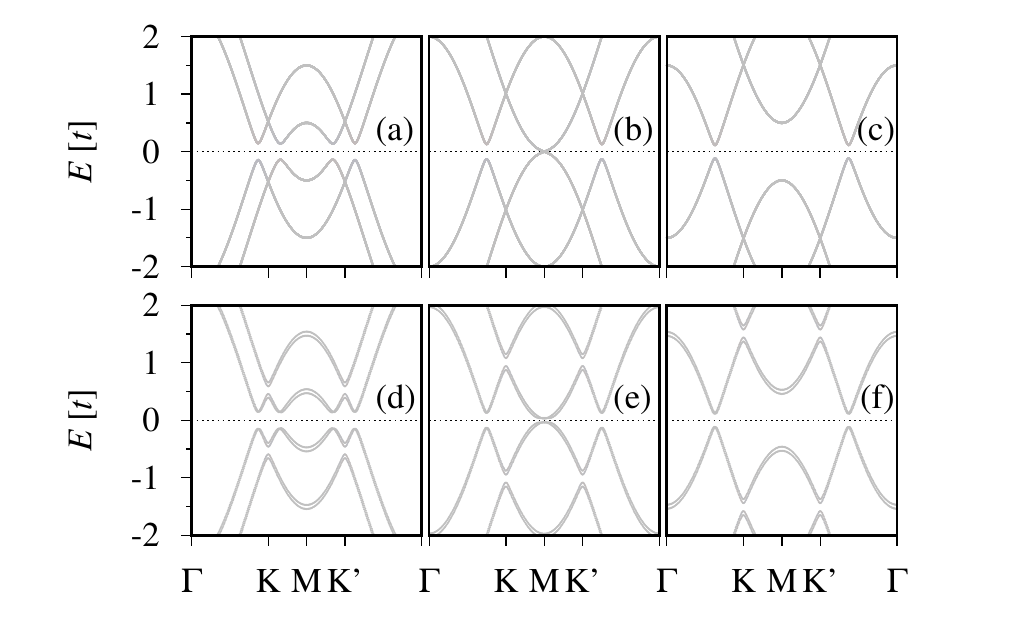}
		\caption{Calculated bulk band structure along $k_y=0$ for $\Delta_S^A=\Delta_S^B=0$ in (a) with $\lambda_E^A=\lambda_E^B=0.5t$, (b) with $\lambda_E^A=\lambda_E^B=1t$, and (c) with $\lambda_E^A=\lambda_E^B=1.5t$. (d)-(f) The same as in (a)-(c) for $\Delta_S^A=0.07t$ and $\Delta_S^B=0.14t$.  We use $\lambda^A_I=\lambda^B_I=3\sqrt{3}\cdot 0.03t$ and $\lambda_{PIA}=0.045t$ with in-plane magnetization along $\mathbf{\hat{m}}=\left(\sqrt{3}/2,1/2,0\right)$.}
		\label{fig:bulk_evolution_uni_ip}
	\end{figure}
	\subsection{Quantum anomalous Hall effect from in-plane magnetization}
	We first investigate the normal system without superconductivity to find QAHE phases. We compute bulk band gap and Chern numbers from the Bloch Hamiltonian $H_e(\mathbf{ k})$ without superconductivity and explore the $\lambda_I$-$\lambda_E$ parameter space. The results are shown in Fig.~\ref{fig:gap_ip_su_uu}. The case (su) remains trivial but for (uu) we find a QAHE phase with $C=-1$ which is independent of the amplitude of intrinsic SOC and PIA (see Fig.~\ref{fig:gap_rashba_pia} in Appendix~\ref{appendix_rashba}) as long as both are present. The topological phase transition is controlled by the exchange coupling. We present only results for uniform exchange coupling since for staggered exchange not all degeneracies are removed and we cannot easily calculate Chern numbers.
	
	The QAHE phase has been proposed to exist in low buckled honeycomb lattices.\cite{Ren2016a,Ren2017}~Based on our results for graphene on TMDCs, where sizeable PIA SOC has been found, we propose that it could also be realized by means of van der Waals heterostructures with flat graphene.
	
	\subsection{Bulk band structure and chiral Majorana fermions in zigzag nanoribbons}
	To demonstrate that the nontrivial gap opens at the $\mathrm{M}$-point we show the bulk band structures of the normal system for different values $\lambda_E$ across the phase transition in Figs.~\ref{fig:bulk_evolution_uni_ip}(a)-(c). In (d)-(f) we introduce superconducting proximity pairing. The splittings of the bands are very small due to the small values of the superconducting gap we use, compared to the exchange coupling. The values for the superconducting pairing on $A$ and $B$ sublattice need to be different to have nondegenerate bands. We use $\Delta_S^A=0.07t$ and $\Delta_S^B=0.014t$. The BdG Chern number in (d) is $C_{BdG}=0$ and in (f) we get $C_{BdG}=-2$. Alternatively, the degeneracy of the bands could also be lifted by different values of the exchange coupling on $A$ and $B$.
	
	\begin{figure}[t]
		\centering
		\includegraphics[width=1\columnwidth]{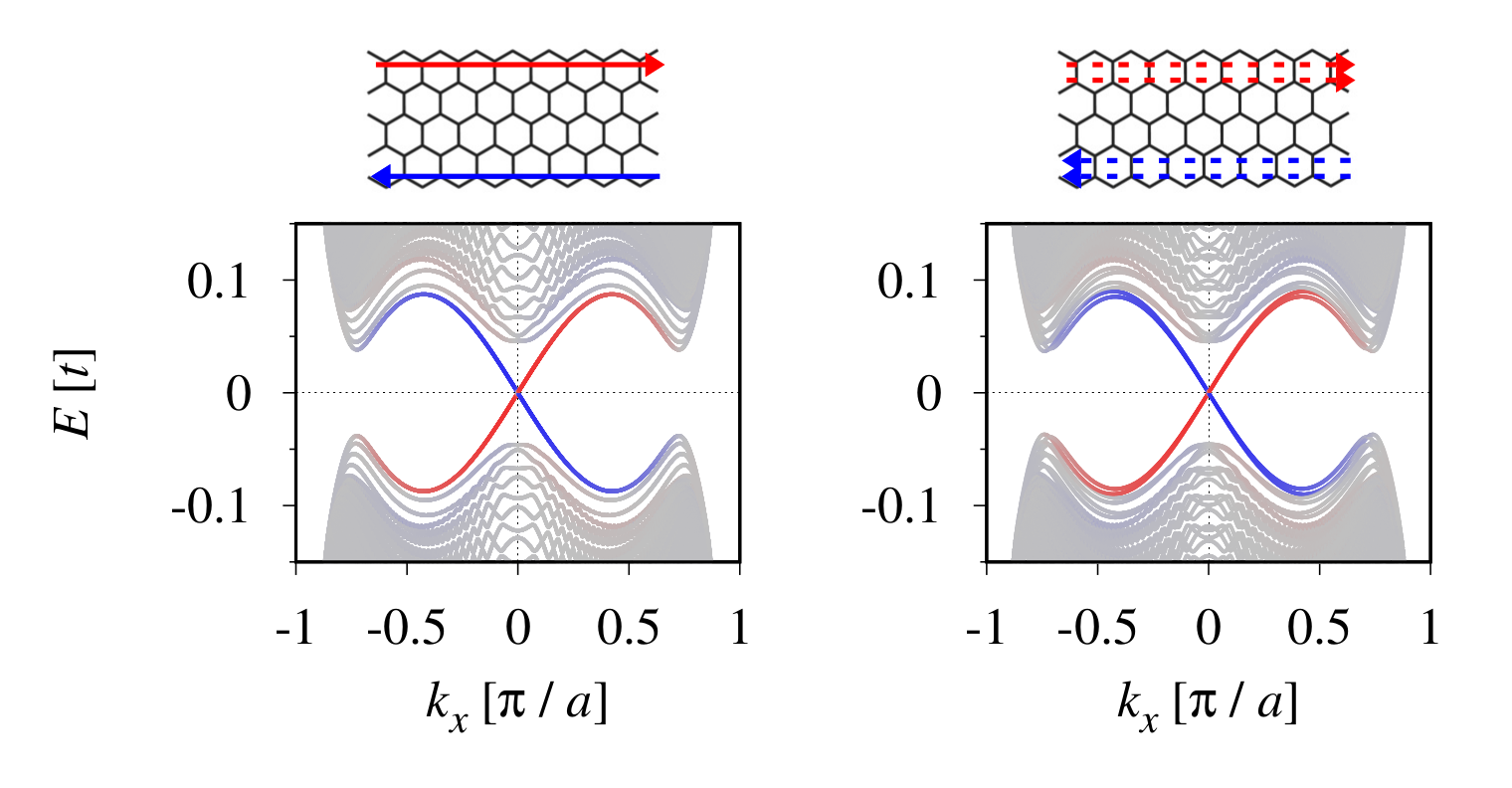}
		\caption{Calculated band structure of zigzag nanoribbons with  width of 100 unit cells for $\Delta_S^A=\Delta_S^B=0$ (left) and $\Delta_S^A=0.07t$, $\Delta_S^B=0.14t$ (right). Color indicates localization of states as shown  in the sketches. Arrows specify direction of propagation for fermionic states (solid line) and Majorana fermions (dashed line). We use $\lambda^A_I=\lambda^B_I=3\sqrt{3}\cdot 0.03t$, $\lambda_{PIA}=0.045t$, and $\lambda_E^A=\lambda_E^B=1.5t$ with in-plane magnetization along $\mathbf{\hat{m}}=\left(\sqrt{3}/2,1/2,0\right)$.}
		\label{fig:ribbon_norm_sc_comp}
	\end{figure}
	Further, we calculate the band structures for zigzag nanoribbons for the QAHE and chiral topological superconductor phases. The results are presented in Fig.~\ref{fig:ribbon_norm_sc_comp}. As expected we find one chiral fermionic state per edge in the QAHE system and two chiral Majorana fermions per edge in the topological superconductor.
	
	\subsection{Phase diagrams and zigzag nanoribbons with single chiral Majorana fermion}
\begin{figure}[t]
		\centering
		\includegraphics[width=1\columnwidth]{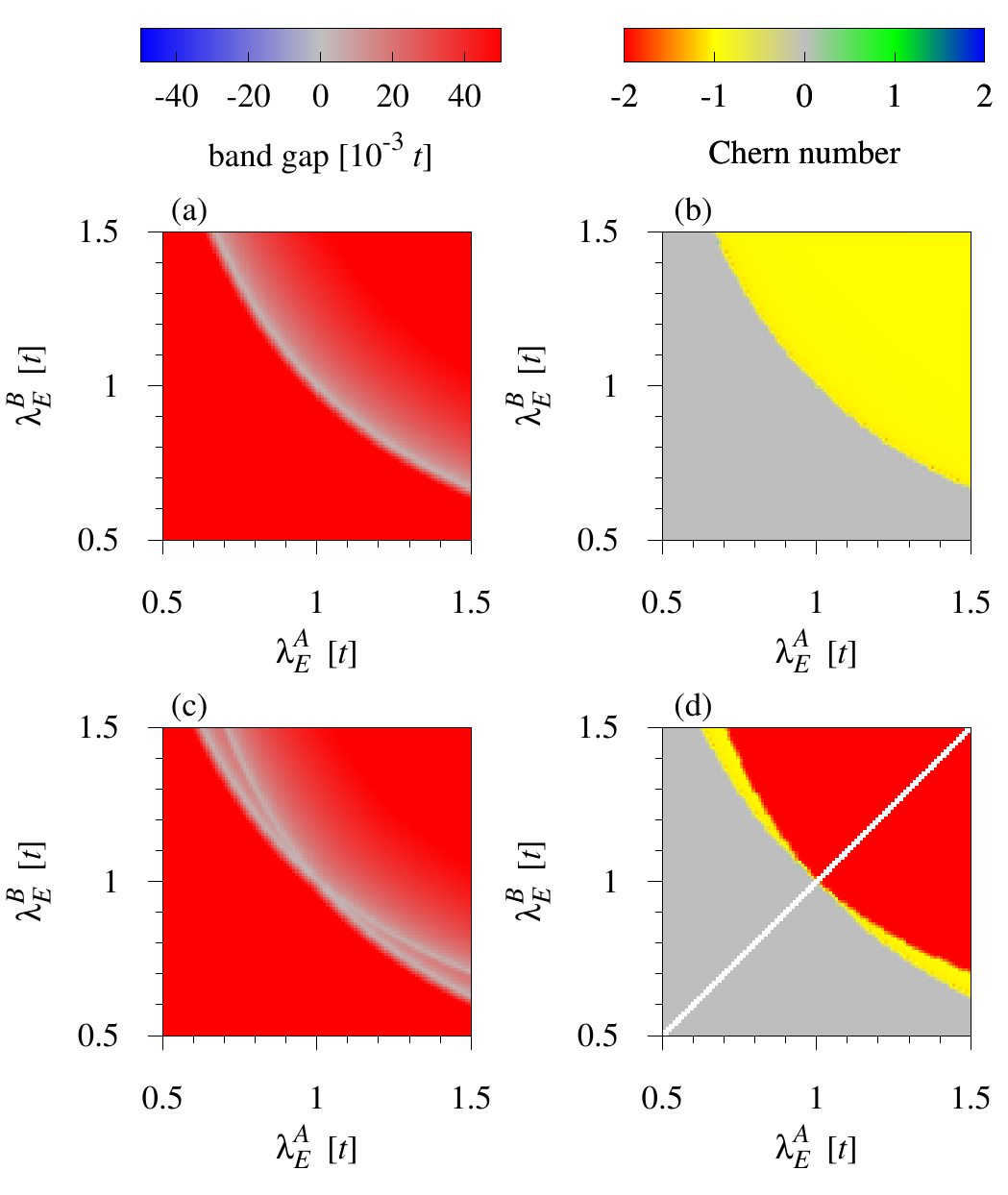}
		\caption{Global bulk band gap (left) and (BdG) Chern number (right). (a) and (b) show a scan of the $\lambda_E^A$-$\lambda_E^B$ parameter space of the QAHE phase diagram ($\Delta_S^A=\Delta_S^B=0$). (c) and (d) show a scan of the the same parameter space of the topological superconductor phase diagram with $\Delta_S^A=\Delta_S^B=0.07t$. We use $\lambda^A_I=\lambda^B_I=3\sqrt{3}\cdot 0.03t$, and $\lambda_{PIA}=0.045t$ with in-plane magnetization along 
			$\mathbf{\hat{m}}=\left(\sqrt{3}/2,1/2,0\right)$.}
		\label{fig:gap_Delta_AB_lE_AB}
	\end{figure}

	Can we now reach a $C_{BdG}=\pm1$ phase by tuning the parameters in the topological superconductor? This can be achieved by introducing an asymmetry to the system such that the phase transitions of the two copies of QAHE for quasiparticles happen at different points in the phase space, transitioning one copy already to a topologically trivial phase while the other one remains nontrivial. The asymmetry can be introduced by breaking sublattice symmetry, i.e., using parameters with different values on $A$ and $B$ sublattice. We show this by scanning the $\lambda_E^A$-$\lambda_E^B$ parameter space (see Fig.~\ref{fig:gap_Delta_AB_lE_AB}). Without superconductivity, we find a topologically trivial phase ($C=0$) and a QAHE phase ($C=-1$). The two regions are separated by a curve along which the bulk band gap closes [Fig.~\ref{fig:gap_Delta_AB_lE_AB}(a) and (b)]. In the presence of superconductivity, the curve splits into two lines when going away from the point $\lambda_E^A=\lambda_E^B=1t$ inducing two new regions to the phase diagram with $C_{BdG}=-1$ between these lines. In this case, the direct transition from $C_{BdG}=-2$ to $C_{BdG}=0$ at the point $\lambda_E^A=\lambda_E^B=1t$ goes through a phase with $C_{BdG}=-1$ when $\lambda_E^A\neq\lambda_E^B$, where only one of the two QAHE copies has entered a trivial phase.
	
\begin{figure}[t]
		\centering
		\includegraphics[width=1\columnwidth]{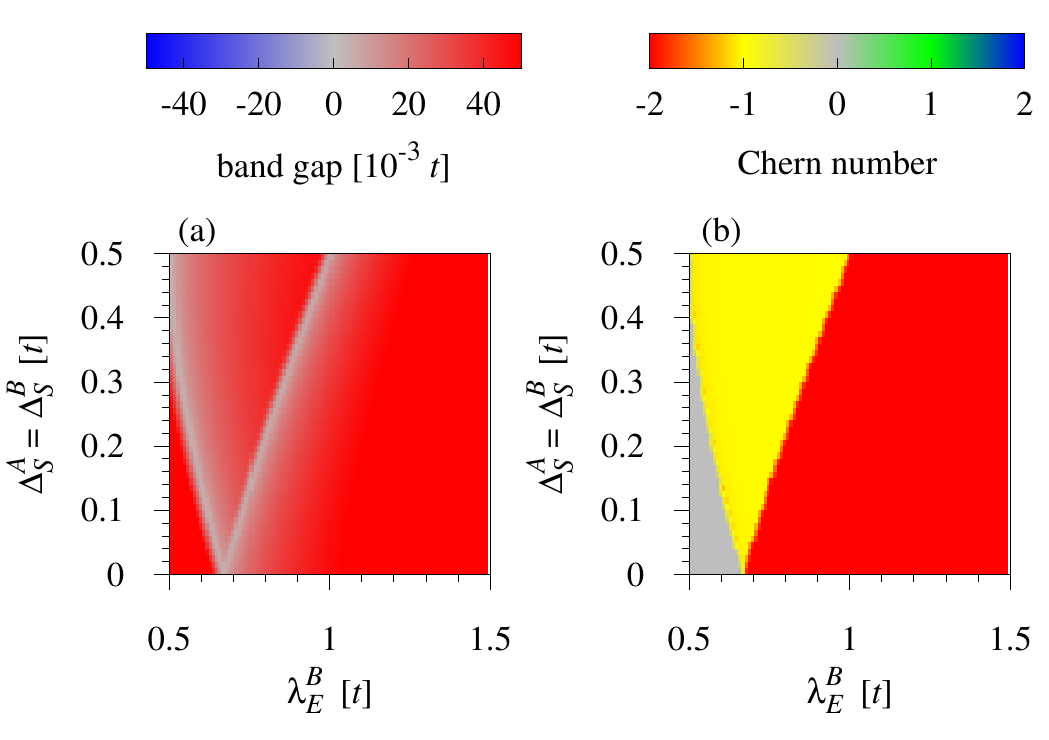}
		\caption{Global bulk band gap (a) and BdG Chern number (b). Along the $x$-axis $\Delta_S^A=\Delta_S^B=0$ showing the QAHE phase transition for fixed $\lambda_E^A=1.5t$ and varying $\lambda_E^B$. Going along the $y$-axis shows the phase space of the superconducting system. We use $\lambda^A_I=\lambda^B_I=3\sqrt{3}\cdot 0.03t$, and $\lambda_{PIA}=0.045t$ with in-plane magnetization along 
			$\mathbf{\hat{m}}=\left(\sqrt{3}/2,1/2,0\right)$.}
		\label{fig:gap_pia_chiral_lEA1_5_lEB_delta}
	\end{figure}

	Close to the region, where the phase transition happens in the non-superconducting system, topological phases with $C_{BdG}=-1$ can be introduced in the analogous superconducting system. We demonstrate this in Fig.~\ref{fig:gap_pia_chiral_lEA1_5_lEB_delta}. Going along the $x$-axis shows the transition from a trivial insulator to a QAHE phase when tuning the exchange coupling on sublattice $B$ while keeping it fixed on $A$. The transition appears at a point where the bulk band gap closes. When the system is turned into a superconductor (going along the $y$-axis), the gap closing point splits into two subsequent gap closings that go further apart from each other with increasing superconducting coupling. In the superconducting phase diagram, this leads to a new phase with $C_{BdG}=-1$ which appears between the trivial ($C_{BdG}=0$) and the topological superconducting phase with even BdG Chern number $C_{BdG}=-2$. Corresponding bulk band structures at distinct points in this phase diagram showing the evolution of the bulk bands across the phase transitions are presented in Appendix~\ref{singleMFevolution}. 
	
	The phase space for sublattice asymmetric superconducting pairing and uniform exchange coupling is presented in Appendix~\ref{appendix_Delta}, where we show that different superconducting coupling on the sublattices can also lead to a topological superconductor with $C_{BdG}=-1$.
	
\begin{figure}[t]
		\centering
		\includegraphics[width=1\columnwidth]{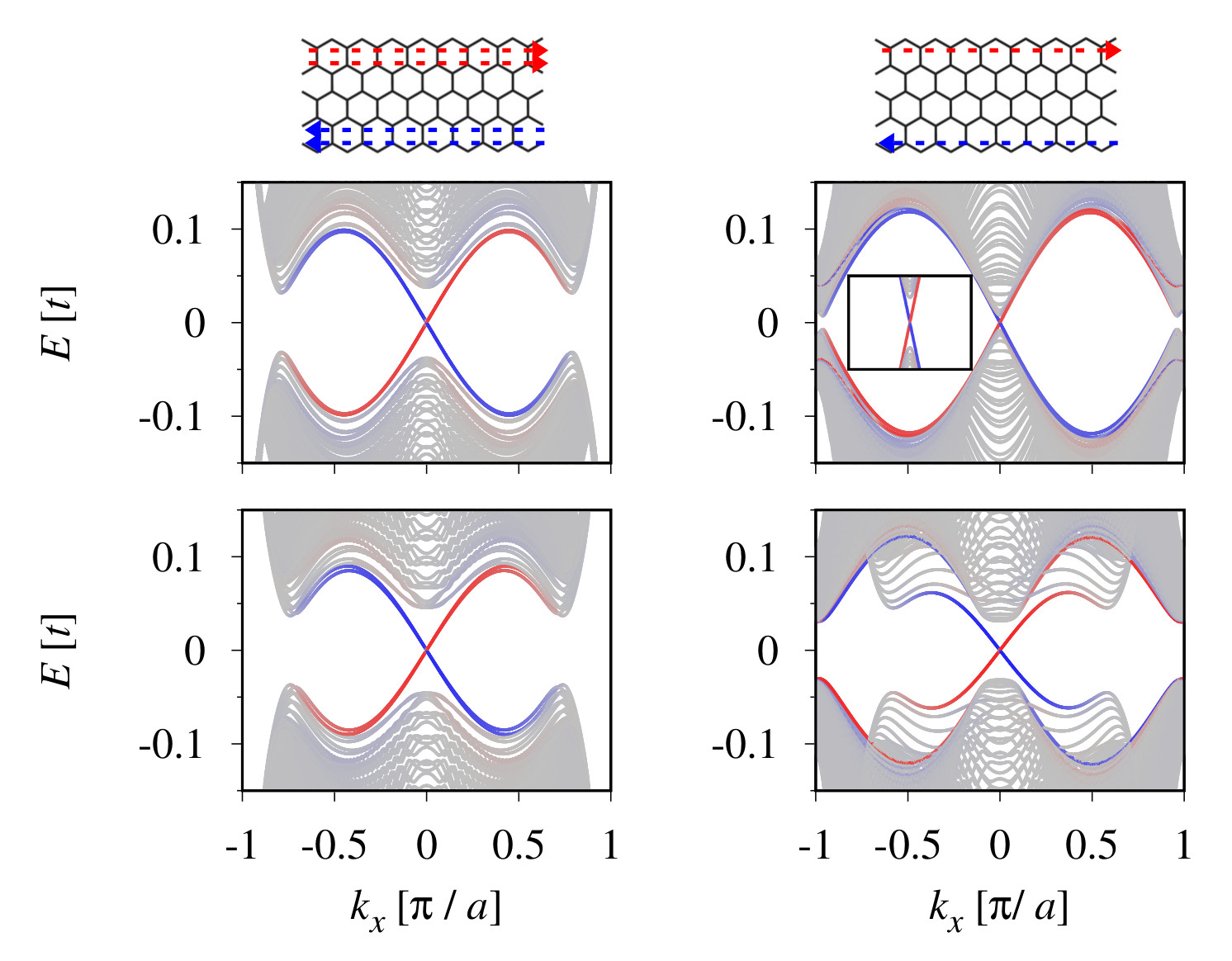}
		\caption{Calculated band structure of zigzag nanoribbons with a width of 100 unit cells. Top panels show spectra for $\Delta_S^A=\Delta_S^B=0.07t$ with $\lambda_E^A=1.5t$, $\lambda_E^B=1.2t$ (left, $C_{BdG}=-2$) and $\lambda_E^A=1.5t$, $\lambda_E^B=0.65t$ (right, $C_{BdG}=-1$). The inset shows a zoom to the edge states around $k_x=0$. Color indicates localization of states as indicated in the sketches. Bottom panels show spectra for $\lambda_E^A=\lambda_E^B=1.5t$ with $\Delta_S^A=0.07t$, $\Delta_S^B=0.014t$ (left, $C_{BdG}=-2$) and $\Delta_S^A=0.07t$, $\Delta_S^B=0.89t$ (right, $C_{BdG}=-1$). The dashed arrows specify the direction of propagation of Majorana fermions. We use $\lambda^A_I=\lambda^B_I=3\sqrt{3}\cdot 0.03t$ and $\lambda_{PIA}=0.045t$ with in-plane magnetization along $\mathbf{\hat{m}}=\left(\sqrt{3}/2,1/2,0\right)$.}
		\label{fig:ribbon_sc_comp_2_1_MF}
	\end{figure}

	Finally, we explore zigzag nanoribbon spectra at distinct points of the phase diagrams [from Fig.~\ref{fig:gap_Delta_AB_lE_AB}(d) and Fig.~\ref{fig:gap_Delta_AB}(b)] with $C_{BdG}=-2$ and $C_{BdG}=-1$ in Fig.~\ref{fig:ribbon_sc_comp_2_1_MF}. When the system is in a single Majorana phase one pair of chiral edge states vanishes. Inside the gap of this trivial insulator, the chiral Majorana fermions of the still nontrivial part of the quasiparticle-QAHE copy reside. For the asymmetric exchange coupling the states of the trivial system are spectrally quite close to the chiral Majorana edge states, whereas for the asymmetric superconducting pairing the gap is large. Nevertheless, the localization strength is comparable for both cases.
	
	Before we conclude, we briefly discuss superconducting proximity effect in graphene. This effect was first demonstrated using Nb contacts.\cite{Heersche2013}~By now, the family of 2D crystals also comprises superconducting materials that can be used to proximity induce superconductivity in van der Waals heterostructures. Gated monolayer TMDCs, such as MoS$_2$, NbSe$_2$, WS$_2$, or WTe$_2$, show Ising superconductivity that can resist even large magnetic fields.\cite{Lu2015,Xi2015,Xi2016,Lu2018,Sajadi2018}~Strained TMDCs are proposed to exhibit superconducting states\cite{Deng2019}~and layered superconducting materials could serve as possible thin film proximity substrates.\cite{Goto2017}~The search for new 2D superconducting materials, their characterization and description is a very active field of research. While our calculations above are for a model system with assumed different superconducting band gaps (see also Appendix C), the variety of superconducting van der Waals heterostructures gives us confidence in an experimental realization of the predicted effects. At the current stage of research it is still too early (even for DFT-based theory) to pinpoint the specific combination of materials to realize the model and observe nontrivial topological phases with single Majorana fermions in graphene devices. 
	
	\section{Summary}\label{SummaryMF}
	In summary, we study superconducting proximity to QAHE phases in graphene. From analyzing the phase space, we find that chiral topological superconductor phases form with even BdG Chern numbers for out-of-plane magnetization at the $\mathrm{K}$-points. In the zigzag and armchair nanoribbons we show corresponding spectra with chiral states localized at the edges. Inducing sublattice asymmetry to the superconducting pairing does not allow for odd BdG Chern numbers due to the symmetric appearance of gap closings and phase transitions at $\mathrm{K/K'}$. Around the $\mathrm{M}$-point, a QAHE state  with $C=-1$ in the normal system and $C_{BdG}=-2$ with superconducting proximity can be induced with in-plane orientation of magnetization. We show that in this case a transition to $C_{BdG}=-1$ is possible either by sublattice asymmetric exchange coupling or superconducting pairing leading to two subsequent phase transitions, one for each copy of QAHE from the quasiparticles. In between only one copy is topological resulting in $C_{BdG}=-1$, which is the topological phase of interest to achieve non-Abelian statistics.	
	
	\section{Acknowledgments}
	This work has been funded by the Deutsche Forschungsgemeinschaft (DFG, German Research Foundation) – Project-ID 314695032 – SFB 1277 and DFG SPP 1666, EU Seventh Framework Programme 
	under Grant Agreement No. 604391 Graphene Flagship, VVGS-2018-1227, and VEGA 1/0105/20.
	
	\appendix

\section{Quantum anomalous Hall effect phase space analysis for further system parameters}\label{appendix_rashba}
	\begin{figure}[h!]
	\centering
	\includegraphics[width=1\columnwidth]{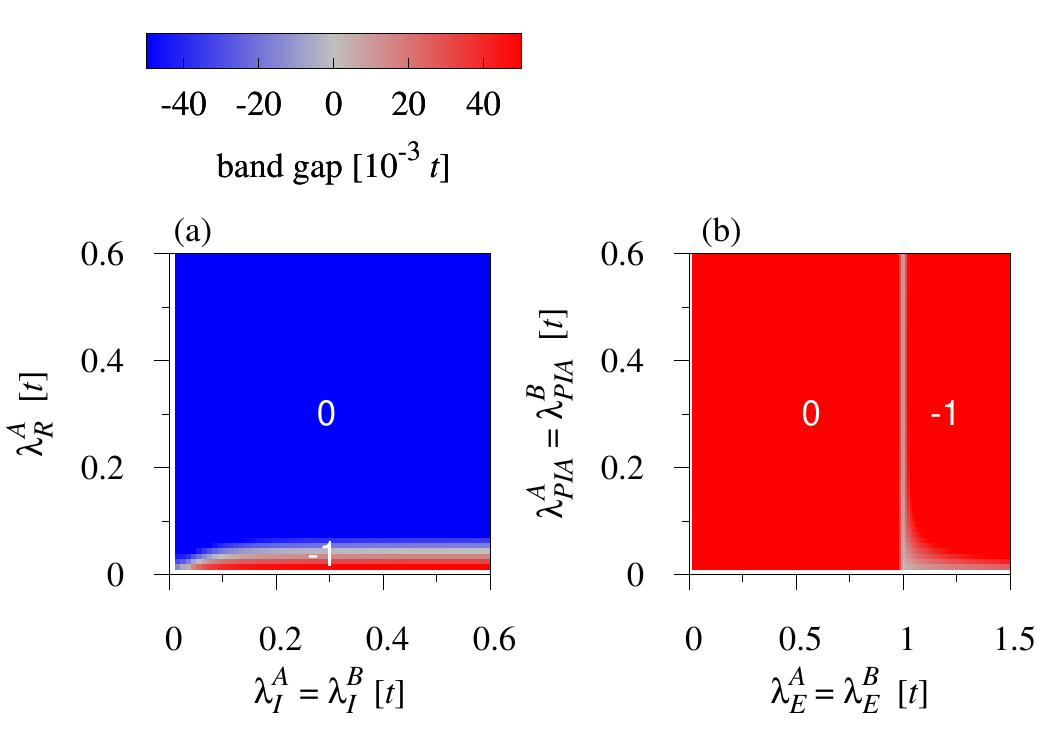}
	\caption{Global bulk band gap and Chern number (white numbers) for (a) $\lambda_{PIA}^A=\lambda_{PIA}^B=0.045t$ and $\lambda_{E}^A=\lambda_{E}^B=1.5t$ with varying intrinsic SOC $\lambda^A_I$, $\lambda^B_I$ and Rashba SOC $\lambda_R$ for in-plane magnetization along 
		$\mathbf{\hat{m}}=\left(\sqrt{3}/2,1/2,0\right)$. (b) shows the same for $\lambda^A_I=\lambda^B_I=3\sqrt{3}\cdot 0.03t$ and $\lambda_R=0$ with varying exchange coupling $\lambda^A_E$, $\lambda^B_E$ and PIA SOC $\lambda_{PIA}^A$, $\lambda_{PIA}^B$.}
	\label{fig:gap_rashba_pia}
\end{figure}
In the main text we have set Rashba SOC to zero since a too large value destroys the topological phase. In Fig.~\ref{fig:gap_rashba_pia}(a) we show that the topologically nontrivial phase survives up to $\lambda_R\approx 0.05t$.
	
	The presence of both, intrinsic SOC and PIA SOC, is necessary because they guarantee the presence of a bulk band gap even along high symmetry lines in the Brillouin zone. Their strength has no direct influence on the phase transitions [see (a) for intrinsic SOC and (b) for PIA SOC] but they need to have a finite value.
	
	The phase transition is directly controlled by the exchange coupling, as shown in Fig.~\ref{fig:gap_rashba_pia}(b). Its orientation is chosen along $\mathbf{\hat{m}}=\left(\sqrt{3}/2,1/2,0\right)$ in the plane in order to have nondegenerate bands and a bulk band gap.~\cite{Ren2016a}
	
	\section{Evolution of bulk band structure along phase transition: trivial, $C_{BdG}=-1$, $C_{BdG}=-2$}\label{singleMFevolution}
	\begin{figure}[t]
		\centering
		\includegraphics[width=1.1\columnwidth]{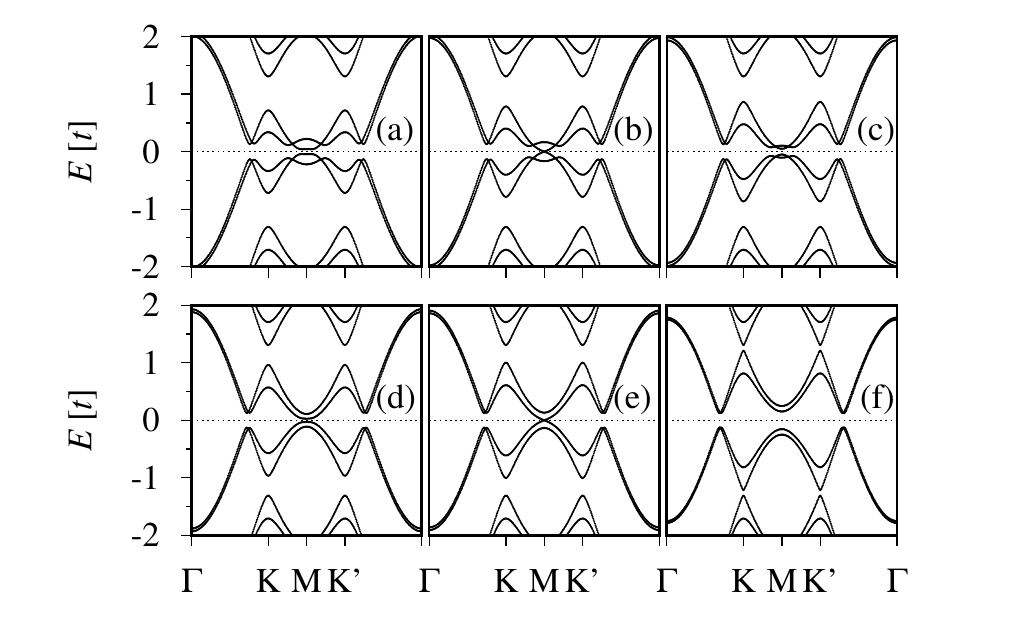}
		\caption{Evolution of bulk bands along phase transition: trivial - single chiral Majorana fermion - pair of chiral Majorana fermions. (a) Trivial superconductor $C_{BdG}=0$ for $\lambda_E^B=0.5t$. (b) First gap closing at $\lambda_E^B=0.57t$. (c) Topological superconductor $C_{BdG}=-1$ for $\lambda_E^B=0.65t$. (d) Topological superconductor $C_{BdG}=-1$ for $\lambda_E^B=0.75t$. (e) Second gap closing at $\lambda_E^B=0.79t$. (f) Topological superconductor $C_{BdG}=-2$ for $\lambda_E^B=1.0t$. We use parameters $\lambda_E^A=1.5t$, $\Delta_S^A=\Delta_S^B=0.2t$, $\lambda^A_I=\lambda^B_I=3\sqrt{3}\cdot 0.03t$, and $\lambda_{PIA}=0.045t$ with in-plane magnetization along 
			$\mathbf{\hat{m}}=\left(\sqrt{3}/2,1/2,0\right)$.}
		\label{fig:bands_singleMF}
	\end{figure}
	We show the corresponding bulk band structures of distinct points in the phase space shown in Fig.~\ref{fig:gap_pia_chiral_lEA1_5_lEB_delta} for a superconducting system with $\Delta_S^A=\Delta_S^B=0.2t$. Going along a line parallel to the $x$-axis in the phase diagram in Fig.~\ref{fig:gap_pia_chiral_lEA1_5_lEB_delta}, i.e. increasing $\lambda_E^B$ while keeping the superconducting coupling fixed, the system is first in a trivial superconducting phase with gapped bulk bands as presented in Fig.~\ref{fig:bands_singleMF}(a). By enhancing the value of the exchange coupling on sublattice $B$, the bulk band gap is then closed [see (b)]. One pair of the four low-energy bands, which is connected via particle-hole symmetry, touches. Increasing the exchange coupling further reopens the gap [see (c)] and the system enters a topologically nontrivial phase with BdG Chern number $C_{BdG}=-1$. The splitting between this pair of particle-hole bands keeps increasing with growing $\lambda_E^B$, while the gap between the other particle-hole symmetric pair of low-energy bands decreases [see (d)] and closes [see (e)]. This leads to a second phase transition to a chiral topological superconductor with $C_{BdG}=-2$ after reopening the bulk gap [see (f)]. The evolution of the bulk bands demonstrates the consecutive phase transitions of the two QAHE copies of the quasiparticles in the chiral topological superconductor.
	
	\section{Single chiral Majorana fermion from asymmetric superconducting gap}\label{appendix_Delta}
	We find for the $\Delta_S^A$-$\Delta_S^B$ parameter space a gap closing when the asymmetry between the superconducting pairing on $A$ and $B$ is large enough accompanied by a phase transition of one copy of quasiparticle-QAHE indicated by the change in the BdG Chern number from $-2$ to $-1$.
	\begin{figure}[b]
		\centering
		\includegraphics[width=1\columnwidth]{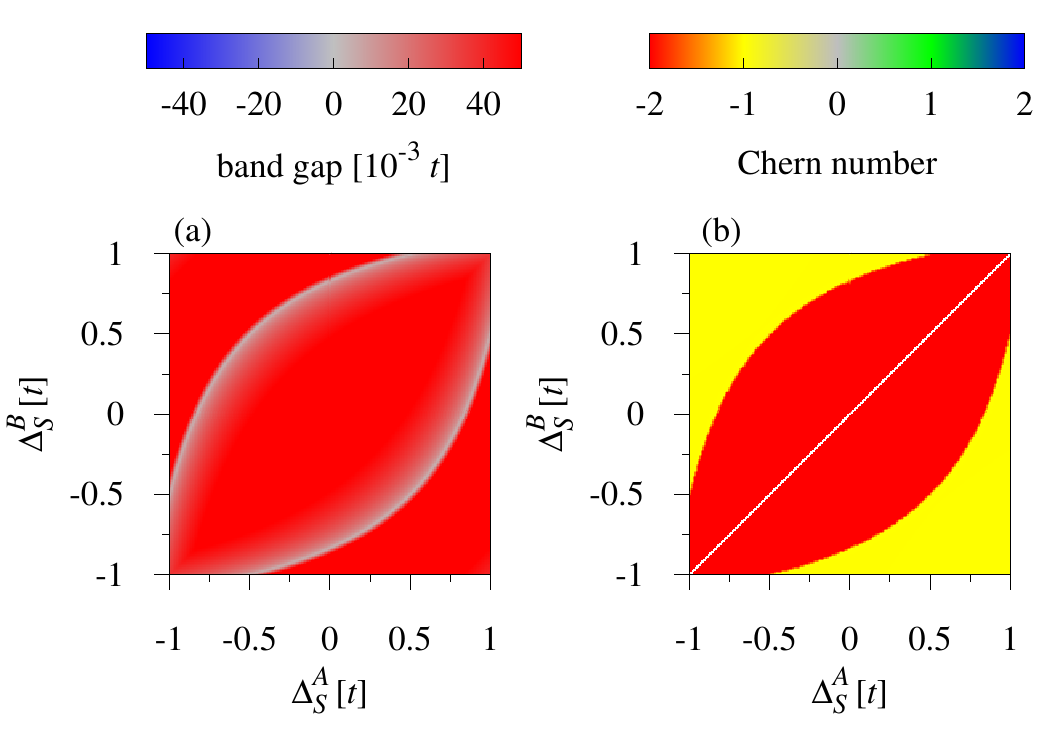}
		\caption{Global bulk band gap (a) and BdG Chern number (b). (a) and (b) show a scan of the $\Delta_S^A$-$\Delta_S^B$ parameter space for $\lambda_E^A=\lambda_E^B=1.5t$. We use $\lambda^A_I=\lambda^B_I=3\sqrt{3}\cdot 0.03t$, and $\lambda_{PIA}=0.045t$ with in-plane magnetization along 
			$\mathbf{\hat{m}}=\left(\sqrt{3}/2,1/2,0\right)$.}
		\label{fig:gap_Delta_AB}
	\end{figure}

	\bibliography{paper}
	
\end{document}